\documentclass[a4paper,12pt]{article}
\usepackage{epsfig}
\usepackage[dvips,usenames]{color}
\usepackage{graphicx}

\newlength{\dinwidth}
\newlength{\dinmargin}
\begin{document}
\title{Effects of the little Higgs Models on Single Top Production at the $e^{-}\gamma$ Colliders }
\bigskip
\author{Yao-Bei Liu$^{1}$, Jie-Fen Shen$^{2}$, Xue-Lei Wang$^{3}$ \\
{\small 1: Henan Institute of Science and Technology, Xinxiang
453003, P.R.China}
\thanks{E-mail:hnxxlyb2000@sina.com}\\
{\small 2: Xinxiang Medical University, Xinxiang  453003, P.R.China}\\
{\small 3: College of Physics and Information
Engineering,}\\
\small{Henan Normal University, Xinxiang  453007, P.R.China}
 }
\maketitle
%\date{today}
\begin{abstract}
\indent  In the framework of the littlest Higgs($LH$) model and the
littlest Higgs model with T-parity($LHT$), we investigate the single
top production process $e^{-}\gamma\rightarrow \nu_{e}b\bar{t}$, and
calculate the corrections of these two models to the cross section
of this process. We find that in the reasonable parameter space, the
correction terms for the tree-level $Wtb$ couplings coming from the
$LHT$ model can generate significantly corrections to the cross
section of this process, which might be detected in the future high
energy linear $e^{+}e^{-}$ collider($ILC$) experiments. However, the
contributions of the new gauge boson $W^{\pm}_{H}$ predicted by the
$LH$ model to this process are very small.
\end{abstract}
PACS number(s): 12.60.Cn, 13.66.Hk, 14.65.Ha\\
\newpage
\indent Recently, the little Higgs model offers a very promising
solution to the hierarchy problem in which the Higgs boson is
naturally light as a result of nonlinearly realized
symmetry\cite{little-1}. The key feature of this model is that the
Higgs boson is a pseudo-Goldstone boson of an approximate global
symmetry which is spontaneously broken by a vacuum expectation
value(VEV) at a scale of a few TeV and thus is naturally light. The
most economical little Higgs model is the so-called
 littlest Higgs model, which is based on a $SU(5)/SO(5)$
 nonlinear sigma model \cite{littlest}. It consists of a $SU(5)$ global
 symmetry, which is spontaneously broken down to $SO(5)$ by a vacuum
 condensate $f$. In this model, a set of new heavy gauge bosons$(B_{H},Z_{H},W_{H})$ and
 a new heavy-vector-like quark(T) are introduced which just cancel
 the quadratic divergence induced by the $SM$ gauge boson loops and the
 top quark loop, respectively. Furthermore, these new particles might produce characteristic signatures
 at the present and future collider experiments
 \cite{signatures-1,signatures-2}.\\
 \indent  It has been shown that the $LH$ model suffers from severe
 constraints from the precision electroweak measurement, which would
 require raising the mass of new particles to be much higher than 1
 TeV\cite{signatures-3}. To avoid this problem, T-parity is
 introduced into the $LH$ model, which is called $LHT$ model\cite{LHT}.
 Under T-parity, the $SM$ particles are T-even and most of the new
 heavy particles are T-odd. Thus, the $SM$ gauge bosons can-not mix
 with the new gauge bosons, and the electroweak precision
 observables are not modified at tree level. In the top-quark
 sector, the $LHT$ model contains a T-odd and T-even partner of the
 top quark. The T-even partner of the top quark cancels the
 quadratic divergence contribution of top quark to Higgs boson mass
 and mixes with top quark. It has been shown that the loop corrections to
 precision electroweak observables are much small and the scale
 parameter parameter $f$ can be decreased to 500GeV\cite{LHT,LHT2}. Thus, this
 model can produce rich phenomenology in the present and future
 experiments.\\
 \indent The top quark is by far the heaviest known fermion with a
 mass of the order of the electroweak scale $m_{t}=172.7\pm2.9GeV$ \cite{topmass}.
 Assuming this is not a coincidence, it is hoped that a detailed
 study of top quark couplings to other particles will be of utility
 in clarifying whether the $SM$ provides the correct mechanism for
 electroweak symmetry-breaking, or whether new physics is
 responsible. It is therefore of interest to provide a general
 description of the top quark couplings, which might be modified due
 to the presence of new interactions or particles. \\
 \indent Future linear colliders are expected to be designed to function also
 as  $\gamma\gamma$ or
 $e\gamma$ colliders with the photon beams generated by laser-scattering
 method, in these modes the flexibility in polarizing both lepton and photon
 beams will allow unique opportunities to analyze the top quark properties and interactions.
  The aim of this
 paper is to consider the process $e^{-}\gamma\rightarrow \nu_{e}b\bar{t}$
 in the context of the $LH$ model and the $LHT$ model, respectively, and
 see whether the effects of these two models on this process can be
 detected in the future $ILC$ experiments.\\
\indent In the $LH$ model, the couplings constants of the $SM$ gauge
boson $W$ and the new heavy gauge boson $W_{H}$ to ordinary
particles, which are related to our calculation, can be written as
\cite{signatures-1}:
\begin{eqnarray}
g_{V}^{We\nu}&=&-g_{A}^{We\nu}=\frac{ie}{2\sqrt{2}s_{W}}[1-\frac{v^{2}}
{2f^{2}}c^{2}(c^{2}-s^{2})],\\
g_{V}^{W_{H}e\nu}&=&-g_{A}^{W_{H}e\nu}=-\frac{ie}{2\sqrt{2}s_{W}}\frac{c}{s},\\
g_{V}^{Wtb}&=&-g_{A}^{Wtb}=\frac{ie}{2\sqrt{2}s_{W}}[1-\frac{v^{2}}
{2f^{2}}(x_{L}^{2}+c^{2}(c^{2}-s^{2}))],\\
g_{V}^{W_{H}tb}&=&-g_{A}^{W_{H}Tb}=-\frac{ie}{2\sqrt{2}s_{W}}\frac{c}{s}
,
\end{eqnarray}
where $f$ is the scalar parameter, $v=246GeV$ is the electroweak
scale, $s_{W}$ represents the sine of the weak mixing angle, and c
is the mixing parameter between $SU(2)_{1}$ and $SU(2)_{2}$ gauge
bosons with $s=\sqrt{1-c^{2}}$. $x_{L}$ is the mixing parameter
between the $SM$ top quark $t$ and the vector-like top quark $T$,
which is defined as
$x_{L}=\lambda_{1}^{2}/(\lambda_{1}^{2}+\lambda_{2}^{2})$,
$\lambda_{1}$ and $\lambda_{2}$ are the Yukawa couplings parameters.
We write the gauge boson-fermion couplings in the form of
  $i\gamma^{\mu}(g_{V}+g_{A}\gamma^{5})$.\\
   \indent  Compared with the process $e^{-}\gamma\rightarrow \nu_{e}b\bar{t}$ in the $SM$,
    this process in the $LH$ model
 receives additional contributions from the heavy boson $W^{\pm}_{H}$ proceed through the Feynman diagrams
 depicted in Fig1. Furthermore, the modification of the relations
 among the $SM$ parameters, the precision electroweak input
 parameters, the correction terms to the $SM$ We$\nu_{e}$ and
 $Wbt$ coupling can also produce corrections to this
 process.\\
\begin{figure}[h]
\begin{center}
\epsfig{file=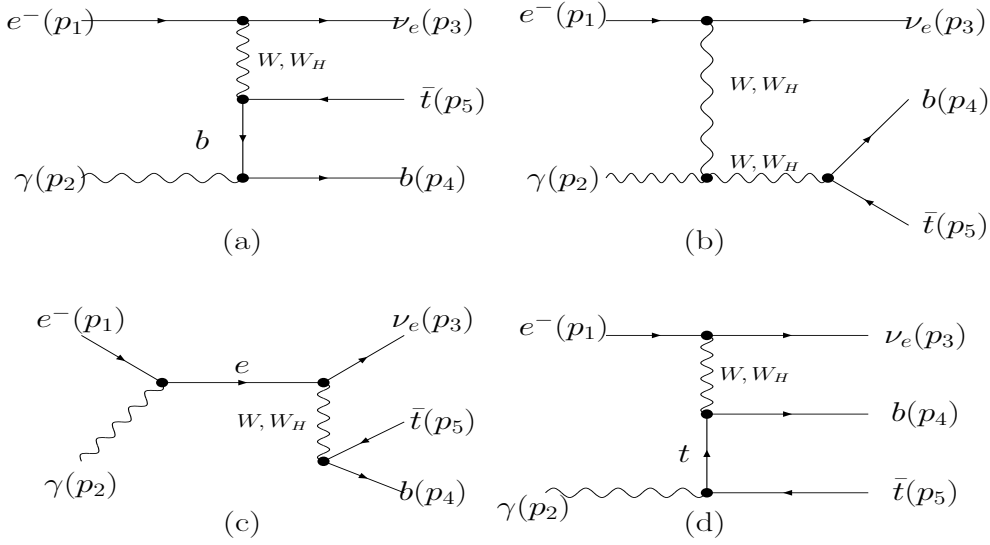,width=450pt,height=500pt} \vspace{-9cm}
\caption{\small Feynman diagrams of the process
$e^{-}\gamma\rightarrow \nu_{e}b\bar{t}$ in the LH model.}
\label{fig1}
\end{center}
\end{figure}
 \indent In order to write a compact expression for the
amplitudes, it is
 necessary to define the triple-boson couplings coefficient as:
\begin{equation}
 \Gamma^{\alpha\beta\gamma}(p_{1},p_{2},p_{3})=g^{\alpha\beta}(p_{1}-p_{2})^{\gamma}
 +g^{\beta\gamma}(p_{2}-p_{3})^{\alpha}+g^{\gamma\alpha}(p_{3}-p_{1})^{\beta},
 \end{equation}
 with all motenta out-going.\\

The invariant production amplitudes of the process in the $LH$ model
 can be written as:
\begin{equation}
 M=M_{a}+M_{b}+M_{c}+M_{d} ,
 \end{equation}
 with
 \begin{eqnarray}
 M_{a}&=&\bar{u}(p_{3})g_{V}^{We\nu}\gamma_{\mu}(1-\gamma_{5})u(p_{1})\{
 G(p_{3}-p_{1},M_{W})+\frac{c^{2}}{s^{2}}G(p_{3}-p_{1},M_{W_{H}})\}\nonumber\\
& &\times g^{\mu\nu}\bar{u}(p_{4})g^{\gamma
b\bar{b}}\gamma_{\rho}G(p_{4}-p_{2},m_{b})
 g_{V}^{Wtb}\gamma_{\nu}(1-\gamma_{5})v(p_{5})\varepsilon^{\rho}(p_{2}),\\
 M_{b}&=&-\bar{u}(p_{3})g_{V}^{We\nu}\gamma_{\mu}(1-\gamma_{5})u(p_{1})
 \Gamma^{\mu\nu\rho}(p_{3}-p_{1},-p_{2},p_{4}+p_{5})\{G(p_{3}-p_{1},M_{W})\nonumber\\
 & &\times
 G(p_{4}+p_{5},M_{W})+\frac{c^{2}}{s^{2}}G(p_{3}-p_{1},M_{W_{H}})G(p_{4}+p_{5},M_{W_{H}})\}\nonumber\\
 & &\bar{u}(p_{4})g_{V}^{Wtb}\gamma_{\nu}(1-\gamma_{5})v(p_{5})\varepsilon^{\rho}(p_{2}),\\
M_{c}&=&\bar{u}(p_{4})g_{V}^{Wtb}\gamma_{\mu}(1-\gamma_{5})v(p_{5})\{
 G(p_{4}+p_{5},M_{W})+\frac{c^{2}}{s^{2}}G(p_{4}+p_{5},M_{W_{H}})\}\nonumber\\
 & &\times g^{\mu\nu}\bar{u}(p_{3})g_{V}^{We\nu}\gamma_{\nu}(1-\gamma_{5})G(p_{1}+p_{2},0)
 g^{\gamma
 e\bar{e}}\gamma_{\rho}u(p_{1})\varepsilon^{\rho}(p_{2}),\\
M_{d}&=&\bar{u}(p_{3})g_{V}^{We\nu}\gamma_{\mu}(1-\gamma_{5})u(p_{1})\{
 G(p_{3}-p_{1},M_{W})+\frac{c^{2}}{s^{2}}G(p_{3}-p_{1},M_{W_{H}})\}\nonumber\\
& &\times
g^{\mu\nu}\bar{u}(p_{4})g_{V}^{Wtb}\gamma_{\nu}(1-\gamma_{5})G(p_{2}-p_{5},m_{t})
 g^{\gamma
t\bar{t}}\gamma_{\rho}v(p_{5})\varepsilon^{\rho}(p_{2}),
  \end{eqnarray}
 where $G(p,m)=1/(p^{2}-m^{2})$ denotes the propagator of the
 particle.\\
\indent The hard photon beam of the $e\gamma$ collider can be
obtained from laser backscattering at the $e^{+}e^{-}$ linear
collider. Let $\hat{s}$ and $s$ be the center-of-mass energies of
the $e\gamma$ and $e^{+}e^{-}$ systems, respectively. After
calculating the cross section $\sigma(\hat{s})$ for the subprocess
$e^{-}\gamma\rightarrow \nu_{e}b\bar{t}$, the total cross section at
the $e^{+}e^{-}$ linear collider can be obtained by folding
$\sigma(\hat{s})$ with the photon distribution function that is
given in Ref\cite{function}:
\begin{equation}
\sigma(tot)=\int^{x_{max}}_{(M_{t}+M_{b})^{2}/s}dx\sigma(\hat{s})f_{\gamma}(x),
 \end{equation}
where
\begin{equation}
f_{\gamma}(x)=\frac{1}{D(\xi)}[1-x+\frac{1}{1-x}-\frac{4x}{\xi(1-x)}+\frac{4x^{2}}{\xi^{2}(1-x)^{2}}],
\end{equation}
with
\begin{equation}
D(\xi)=(1-\frac{4}{\xi}-\frac{8}{\xi^{2}})\ln(1+\xi)+\frac{1}{2}+\frac{8}{\xi}-\frac{1}{2(1+\xi)^{2}}.
\end{equation}
In the above equation, $\xi=4E_{e}\omega_{0}/m_{e}^{2}$ in which
$m_{e}$ and $E_{e}$ stand, respectively, for the incident electron
mass and energy, $\omega_{0}$ stands for the laser photon energy,
and $x=\omega/E_{e}$ stands for the fraction of energy of the
incident electron carried by the backscattered photon. $f_{\gamma}$
vanishes for $x>x_{max}=\omega_{max}/E_{e}=\xi/(1+\xi)$. In order to
avoid the creation of $e^{+}e^{-}$ pairs by the interaction of the
incident and backscattered photons, we require
$\omega_{0}x_{max}\leq m_{e}^{2}/E_{e}$, which implies that $\xi\leq
2+2\sqrt{2}\simeq4.8$. For the choice of $\xi=4.8$, we obtain
\begin{equation}
x_{max}\approx0.83,~~~~~~~~~~~~~~~~~~~D(\xi_{max})\approx1.8.
 \end{equation}
 For simplicity, we have ignored the possible polarization for the
 electron and photon beams.\\
\indent With the above production amplitudes, we can obtain the
production cross section directly. In the calculation of the cross
section, instead of calculating the square of the amplitudes
analytically, we calculate the amplitudes numerically by using the
method of the references \cite{HZ} which can greatly simplify our
calculation. \\
\indent In our numerical results, we take the input parameters as
 $M_{t}=172.7GeV$\cite{topmass}, $\alpha_{e}=1/128.8$, $M_{Z}=91.187GeV$, $s_{W}^{2}$=0.2315 and
 $m_{W}$=80.45GeV\cite{data}. The value of the relative correction
parameter is insensitive to the degree of the electron and positron
polarization and the c.m. energy $\sqrt{s}$. Therefore, we do not
consider the polarization of the initial states and take
$\sqrt{s}$=500GeV in our numerical calculation. Except for these
$SM$ input parameters, the contributions of the $LH$ model to single
top quark production are dependent on the free parameters ($f$, c,
$x_{L}$). Considering the constraints of the electroweak precision
data on these free parameters, we will assume $f=1\sim2TeV$,
$0.3\leq x_{L}\leq0.6$, $0<c\leq0.5$ for the $LH$ model \cite{con1}.
The relative correction of the $LH$ model to the cross section of
single top production is in the expression of the relative
correction parameter $\delta\sigma/\sigma^{SM}$ with $\delta\sigma=|
\sigma^{tot}-\sigma^{SM}|$ and $\sigma^{SM}$ is the tree-level cross
section of $e^{-}\gamma\rightarrow \nu_{e}b\bar{t}$
 production predicted by the $SM$.
  The
numerical results are summarized in Figs.2\\
\begin{figure}[t]
\begin{center}
\scalebox{0.9}{\epsfig{file=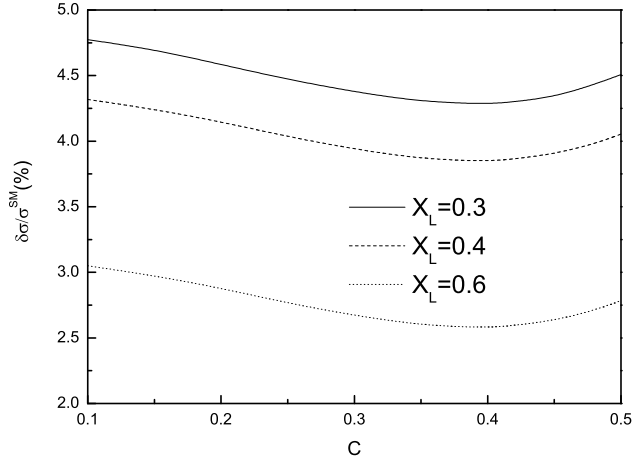}}\\
\end{center}
\caption{\small The relative correction $\delta\sigma/\sigma^{SM}$
as a function of the mixing parameter c for f=1TeV and three values
of the mixing parameter $x_{L}$ in $LH$ model.}
\end{figure}
\indent In the $LH$ model, the extra contributions to the process
$e^{-}\gamma\rightarrow \nu_{e}b\bar{t}$ come from the heavy gauge
boson $W_{H}^{\pm}$, the modification of the relations among the
$SM$ parameters and the precision electroweak input parameters, and
the correction terms of the $SM$ $We\nu_{e}$ and $Wtb$ couplings.
From Fig.2, we can see that the absolute values of the relative
correction $\delta\sigma/\sigma^{SM}$ are smaller than $5\%$ in most
of the parameter space preferred by the electroweak precision data ,
which is difficult to be detected.\\
 \indent  Under T-parity of the $LH$ model, the couplings of
the electroweak gauge boson to light fermions are not modified from
their corresponding $SM$ couplings at tree level. Nonetheless, the
$Wtb$ coupling is modified at tree level by the mixing of the top
quark with its T-even partner\cite{LHT,LHT2}. The expression of the
coupling $Wtb$ can be written as:
\begin{eqnarray}
g_{V}^{Wtb}&=&-g_{A}^{Wtb}=\frac{ie}{2\sqrt{2}s_{W}}[1-\frac{c^{4}_{\lambda}}{2}\frac{v^{2}}
{f^{2}}],
\end{eqnarray}
where the mixing parameter
$c_{L}=\lambda_{1}/\sqrt{(\lambda_{1}^{2}+\lambda_{2}^{2})}$, in
which $\lambda_{1}$ and $\lambda_{2}$ are the Yukawa couplings
parameters in the $LHT$ model.\\
\begin{figure}[t]
\begin{center}
\scalebox{0.9}{\epsfig{file=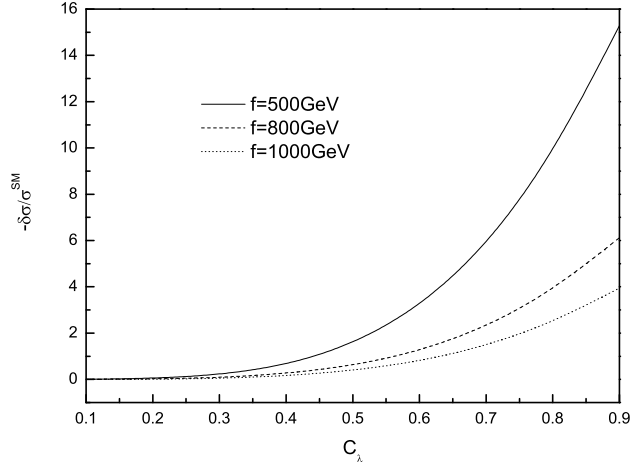}} \caption{\small The relative
correction $\delta\sigma/\sigma^{SM}$ as a function of the mixing
parameters $c_{\lambda}$ for three values of the scale parameter $f$
in $LHT$ model.}
\end{center}
\end{figure}
\begin{figure}[hb]
\begin{center}
\scalebox{0.9}{\epsfig{file=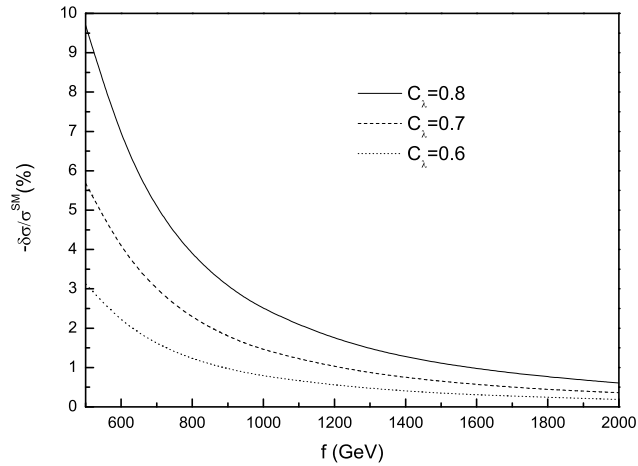}} \caption{\small The relative
correction $\delta\sigma/\sigma^{SM}$ as a function of the the scale
parameter for three values of the mixing parameters $c_{\lambda}$ in
$LHT$ model.}
\end{center}
\end{figure}
\indent From the above discussions, we can see that the $LHT$ model
can also generate corrections to the production cross sections for
the process $e^{-}\gamma\rightarrow \nu_{e}b\bar{t}$ via the
modification of the coupling $Wtb$. The value of the relative
correction parameter $\delta\sigma/\sigma^{SM}$ depends on two free
parameters $f$ and $c_{\lambda}$ in the $LHT$ model. Considering the
parameter space of $f$ and $c_{\lambda}$ constrained by
Ref\cite{LHT2}, we take $0.1\leq c_{\lambda}\leq0.9$ and $500GeV\leq
f\leq2000GeV$. The relative correction parameters generated by the
$LHT$ model to the cross section of single top production at the
$e\gamma$ collider are shown in Fig.3 and Fig.4. In these figures,
we have taken $\delta\sigma=\sigma^{LHT}-\sigma^{SM}$. From Fig.3,
we can see that the absolute value of the relative correction
increases with an increase of the mixing parameter $c_{\lambda}$. As
long as $f\leq800GeV$ and $c_{\lambda}\geq0.7$, in sizable regions
of the parameter space in the $LHT$ model, the absolute value of the
relative correction $\delta\sigma/\sigma^{SM}$ is larger than $5\%$,
which might be detected in the future $ILC$ experiments. To see the
effect of varying the scale parameter $f$ on the relative correction
$\delta\sigma/\sigma^{SM}$, we plot $\delta\sigma/\sigma^{SM}$ as a
function of $f$ for three values of the mixing parameter
$c_{\lambda}$ in Fig.4. One can see from Fig.4 that the absolute
value of the relative correction $\delta\sigma/\sigma^{SM}$
decreases as $f$ increase, which is consistent with the conclusions
for the corrections of the $LH$ model and $LHT$ model to other
observables\cite{yue}.\\
\indent The little Higgs model, which can solve the hierarchy
problem, is a promising alternative new physics model. The $LH$
model is one of the simplest and phenomenologically viable models,
which realizes the little Higgs idea. In order to provide a valuable
theoretical instruction to test the little Higgs idea, people have
done a lot of phenomenological work within the context of the little
Higgs models. In this paper, we have considered single top
production process $e^{-}\gamma\rightarrow \nu_{e}b\bar{t}$ in the
$LH$ model and the $LHT$ model. We find that the contribution of the
$LH$ model to this process is very small in most of the parameter
space, which is difficult to be detected in future $ILC$
experiments. However, in sizable regions of the parameter space in
the $LHT$ model, the absolute value of the relative correction
$\delta\sigma/\sigma^{SM}$ is larger than $5\%$, which might be
detected
in the future $ILC$ experiments.\\

\end{document}